# DEVELOPMENT OF CYBERSECURITY SIMULATOR-BASED PLATFORM FOR THE PROTECTION OF CRITICAL INFRASTRUCTURES

*TREO Paper*

Tero Vartiainen, University of Vaasa, Vaasa, Finland, tero.vartiainen@uwasa.fi
Duong Dang, University of Vaasa, Vaasa, Finland, duong.dang@uwasa.fi
Mike Mekkanen, University of Vaasa, Vaasa, Finland, mike.mekkanen@uwasa.fi
Emmanuel Anti, University of Vaasa, Vaasa, Finland, emmanuel.anti@uwasa.fi

## Abstract

*Critical infrastructures or critical national infrastructures (CNI) are interconnected with communication systems, making them vulnerable to cyberattacks. We propose the development of a platform, based on the real-time simulation of cyber-physical systems, allows CNI stakeholders to strengthen CNIs' resilience and security, by using design science and citizen science as research methods. The platform enables the creation of a Digital Twin (DT) and the execution of its functions in real-time. We have initiated the first steps of the platform in the Vaasa Harbor Microgrid (VHM). The lab provides a co-simulation environment that replicates the processes and systems present in VHM, thereby facilitating the simulation of various cyberattack scenarios on VHM. During the development of a cybersecurity simulator-based platform, we anticipate key contributions including the establishment of principles for designing functions of a cybersecurity simulator-based platform, an in-lab platform for studying and co-developing cyber-attack scenarios, and services for other CNI stakeholders.*

## 1    Introduction

Critical infrastructures, also known as critical national infrastructures (CNI), are vital to societal and economic functions and thus require special protection (Gallais and Filiol 2017). These infrastructures include, for example, telecommunications, electric power systems, information technology networks, and data centers (Curt and Tacnet 2018). Disruptions or failures in CNI can have substantial and far-reaching implications for a nation's safety, economy, and society (Singh et al. 2014). Consequently, nations worldwide have implemented protection programs for CNI. Some examples include the European Programme for Critical Infrastructure Protection (EPCIP), the US's critical infrastructure protection program, and Singapore's Protected Areas and Protected Places Act.

However, protecting CNI is complex due to their multifaceted functions and interconnectedness (Gallais and Filiol 2017).  In addition, the integration of cyber-physical systems (CPS) into CNI introduces new risks. CPS, which combine computational and physical capabilities to interact with other devices and humans, often connect to telecommunications networks, making them vulnerable to cyberwarfare, cyber espionage, and other modern threats (Dawson et al. 2021). Scholars have called for focused research on CPS cybersecurity (Kumar et al. 2020). In fact, efforts have been made to secure CPS for CNI protection, ranging from technical solutions (e.g., encryptions, firewalls) to standards (e.g., Risk Analysis Framework, Cyber Security Framework) and education (e.g., Training Programs, Living Labs) (Dang and Vartiainen 2024). These solutions are often built-in, making them difficult to modify once implemented. Testing them in a real physical system in a hazard mode is challenging. In that sense, a real-time simulator (RTS), a computer model that runs at the same rate as the actual physical system, could reduce development costs and time, simulate different cyber-attack scenarios, and propose response scripts.

RTS has been extensively utilized in the energy sector. However, most RTS literature focuses on regulation, technical aspects, or standards, rather than platform-based solutions for protecting CPS (Dang and Vartiainen 2024). The lack of study in this area can be attributed to the multidisciplinary nature of platform development, which involves fields such as information systems, electrical engineering, and computer science. This TREO talk addresses this gap by presenting our development of a cybersecurity simulator-based platform for the protection of critical infrastructures.

## 2   Background

### 2.1   Critical infrastructures

CNI are vital for a nation's welfare and are vulnerable to various risks. CNIs possess characteristics such as physical, cyber, geographical, and logical dependencies. These dependencies can result in shared vulnerabilities and threats, thereby increasing the complexity and vulnerability of critical infrastructures (Singh et al. 2014). The escalating risk of cyber-attacks has introduced a new dimension to critical infrastructure protection (Dawson et al. 2021; Gallais and Filiol 2017; Singh et al. 2014). CNIs are interconnected with CPS and are categorized into three tiers: user tier (e.g., individuals, businesses), cyber tier (e.g., servers, data centers), and physical tier (e.g., manufacturing, smart grids) (Kumar et al. 2020). These tiers are interconnected via IT systems, which are vulnerable to cyber-attacks such as DoS attack. Simulating CPS or their components can help identify attack vectors, prevent attacks, understand the effects of attacks, and design components to recover from inevitable attacks.

### 2.2   Real-time simulation Digital Twin

A RTS digital twin (DT) understands as a real-time digital replica of a physical device (VanDerHorn and Mahadevan 2021). DTs are utilized to predict the behavior of physical systems under various modeled circumstances, employing the concept of 'what ifs'. This involves the use of digital technologies for sensing, monitoring, diagnosing problems, and optimizing device parameters. Consequently, DTs provide vital information about potential vulnerabilities, enabling swift responses to neutralize attacks and minimize further damage. They also facilitate the acceleration of normal operations restoration, learning from incidents to prevent future recurrences. In addition, DTs enable seamless connection and data exchange with real CNIs using a Hardware-in-the-Loop (HIL) approach. This, in turn, allows for the testing of real physical CNIs, applications, and human behavior using a human-in-the-loop approach (Mekkanen et al. 2022).

## 3   Research Settings and Methods

To address our research gap - the development of a cybersecurity simulator-based platform for the protection of critical infrastructures - we established a laboratory environment within the context of a smart grid. The physical environment is the Vaasa Harbor Microgrid (VHM) located in Vaasa, Finland. Operation data from the VHM is transferred to the Future Reliable Electricity and Energy System Integration Laboratory (FREESI) at the University of Vaasa. Here, the OP5707 XG real-time simulator, a component of FREESI, enables HIL simulations of Smart Grids. The lab provides a co-simulation environment that mimics the processes and systems existing in the VHM, allowing for the simulation of various cyberattack scenarios on the VHM. This provides a realistic environment for platform development. Based on these devices and their capabilities, a digital platform offering a wide range of services based on simulators, covering three tiers: user, cyber, and physical, is developed. The research approach employed in this study is design science (Hevner et al. 2004). This methodology facilitates the development and evaluation of designed artifacts, which, in this context, are the layers of the platform. It also aids in enhancing the functional performance of the artifact. We also implement the strategy of citizen science (Dang et al. 2022) as a vehicle to gather public feedback, which is vital in the development of our platform.

## 4    Development of cybersecurity simulator-based platform

During the development process of a cybersecurity simulator-based platform, several key contributions are anticipated. First, an in-lab cybersecurity simulator-based platform is designed and established. This platform provides services that enable users (e.g., students or corporate partners or other stakeholders) to establish a virtual connection to the platform. This connection facilitates the study and development of cyber attack scenarios and services, and allows users to test the outcomes of their scenarios, using our infrastructure and resources, thereby aiding in their preparedness for potential cyber attacks. Second, we extend various platform functions such as design assistance for scenario modeling, collaboration support for co-development of cybersecurity and resilience, service templates for design tasks, and expert advice for the implementation of simulation software and hardware. Third, we also provide principles as a research contribution, which guide the design of each of our platform functions.

## References

Curt, C., and Tacnet, J.-M. 2018. "Resilience of Critical Infrastructures: Review and Analysis of Current Approaches," *Risk Analysis* 38 (11), 2441–2458.

Dang, D., Mäenpää, T., Mäkipää, J.-P., and Pasanen, T. 2022. "The Anatomy of Citizen Science Projects in Information Systems," *First Monday* 57 (10).

Dang, D., and Vartiainen, T. 2024. "Exploring Socio-Technical Gaps in the Cybersecurity of Energy Informatics for Sustainability," in *Adoption of Emerging Information and Communication Technology for Sustainability*, UK: CRC Press.

Dawson, M., Bacius, R., Gouveia, L. B., and Vassilakos, A. 2021. "Understanding the Challenge of Cybersecurity in Critical Infrastructure Sectors," *Land Forces Academy Review* (26:1), pp. 69–75.

Gallais, C., and Filiol, E. 2017. "Critical Infrastructure: Where Do We Stand Today? A Comprehensive and Comparative Study of the Definitions of a Critical Infrastructure," *Journal of Information Warfare* 16 (1), 64–87.

Hevner, A. R., March, S. T., Park, J., and Ram, S. 2004. "Design Science in Information Systems Research," *MIS Quarterly* 28 (1), 75–105.

Kumar, C., Marston, S., and Sen, R. 2020. "Cyber-Physical Systems (CPS) Security: State of the Art and Research Opportunities for Information Systems Academics," *Communications of the Association for Information Systems* 47 (1).

Mekkanen, M., Vartiainen, T., and Dang, D. 2022. "Develop a Cyber Physical Security Platform for Supporting Security Countermeasure for Digital Energy System," *Scandinavian Simulation Society*, 219–225.

Singh, A. N., Gupta, M. P., and Ojha, A. 2014. "Identifying Critical Infrastructure Sectors and Their Dependencies: An Indian Scenario," *International Journal of Critical Infrastructure Protection* 7 (2), 71–85.

VanDerHorn, E., and Mahadevan, S. 2021. "Digital Twin: Generalization, Characterization and Implementation," *Decision Support Systems* (145), p. 113524.